\documentclass[useAMS,usenatbib]{mn2e}
\usepackage{graphicx} 
\usepackage{subfigure} 
\usepackage{aas_macros}

\newcommand{\msun}{M_{\odot }}
\newcommand{\degree}{\ensuremath{^\circ}}
\title[The instability of planetary systems 
in binaries]{The instability of planetary 
systems in binaries: how the Kozai mechanism 
leads to strong planet-planet interactions}
\author[D. Malmberg, M. B. Davies and J. E. Chambers]
{Daniel Malmberg$^{1}$\thanks{E-mail:danielm@astro.lu.se 
(DM); mbd@astro.lu.se (MBD); chambers@dtm.ciw.edu 
(JC)}, Melvyn B. Davies$^{1},$ John E. 
Chambers$^{2}$  \\ 
$^{1}$Lund Observatory, Box 43, SE--221 00, Lund, 
Sweden \\
$^{2}$Department of Terrestrial Magnetism, 
Carnegie Institution of Washington, \\
5241 Broad Branch Road NW, Washington 
DC 20015, USA}

\begin{document}
\date{Accepted for publication in MNRAS Letters}
\pagerange{\pageref{firstpage}--\pageref{lastpage}} 
\pubyear{2006}
\maketitle
\label{firstpage}

\begin{abstract}
In this letter we consider the evolution of a 
planetary system around a star inside a wide binary.  
We simulate numerically the evolution of the planetary 
orbits for both co-planar and highly-inclined systems.
We find that the Kozai mechanism operates in the 
latter case. This produces a highly eccentric
outer planet whose orbit crosses those of some
of the inner planets. Strong planet-planet 
interactions then follow
resulting in the ejection of one or more planets.
We note that planetary systems resembling our 
solar system, formed around single stars in stellar 
clusters may exchange into binaries and thus
will be vulnerable to planet stripping. This process will 
reduce the number of solar-system like planetary systems,
and may produce at least some of the observed extra-solar
planets.
\end{abstract}


\section{Introduction} \label{sec:intro}
Of the 200 known extrasolar planets
\citep[see for example][]{2006ApJ...646..505B} 
around 40 \citep{2006astro.ph.10623D} 
are in known binary systems, while 
the fraction with an unseen companion (eg. 
brown dwarf) may also be significant 
\citep[see for example][]{2006astro.ph.11542B,
2006astro.ph..9464L}.
A key question is how the presence of the 
companion affects the orbits of the planets and the 
stability of the system. Various effects must
be considered; if the systems consists of one
planet inside a tight binary, a close 
encounter between the planet and the companion
can lead to the ejection of the planet 
\citep{1999AJ....117..621H}.
In wider binaries, secular effects, which
builds up over thousands of orbits are 
important. \citet{1999AJ....117..621H} derived 
limits on the values which the semi-major axis and 
eccentricity of a companion star can have, 
without causing large effects on the orbits of the 
planets. Furthermore, \citet{2005ApJ...618..502M} 
analyzed the importance of planet-planet interactions  
in multi-planet systems with a co-planar 
companion star. If the companion star is highly inclined to 
the orbital plane of the planets, an effect called the 
Kozai Mechanism \citep{1962AJ.....67..591K} becomes 
important. Its importance for 
single-planet systems has been analyzed in for example 
\citet{2005ApJ...627.1001T}. Given that the solar system 
is stable, it is possible to derive limits on the 
properties (mass and distance) that a hypothetical 
companion star to the sun could have, without causing 
severe damage to the solar system 
\citep{1997AJ....113.1915I,1999AJ....117..621H}.
In this letter we extend these studies, 
considering in more detail 
the evolution of planetary systems resembling our own 
solar system, when placed inside a binary. Formation of 
planetary systems in binaries is possible and has been 
studied in for example
\citet{2002A&A...396..219B,2000ApJ...543..328M,
2004A&A...427.1097T,2006Icar..183..193T,
2006ApJ...641.1148B,2006Icar..185....1Q}. 
In this letter however, we only consider  
planet formation around single stars. The key idea 
here is to have such single stars exchange into 
binaries in young stellar clusters. In such three-body
encounters, it is possible that the planetary system will 
be destroyed immediately, but a significant fraction will 
be unaffected, assuming that the final binary
is wide \citep[see for example][]{1994ApJ...424..870D}. 
If this star had a 
planetary system this would now be \textit{randomly} 
oriented with respect to the companion star. It is the 
evolution of these system that we study in 
this letter. In section \ref{sec:mercury} we discuss the 
integrator used to perform the numerical simulations 
presented in this letter. Then we give an overview of the 
Kozai mechanism in section \ref{sec:kozai}. We present the 
results of our numerical simulations in section 
\ref{sec:runs}, which show that planet-planet interactions
can cause the disruption of planetary systems if inside an
inclined binary. In section \ref{sec:discussion} we discuss
the implications for extra-solar planets and the vulnerability
of planetary systems resembling our solar system.

\section{Simulations - the mercury code} \label{sec:mercury}
The numerical simulations in this paper where done using 
a modified version of the publicly available 
\textsc{mercury} code, which we have specifically adapted 
for integrating planetary systems in binary stars. The 
original \textsc{mercury} code is described in 
\citet{1999MNRAS.304..793C}, 
and the modifications in our version are described in 
\citet{2002AJ....123.2884C}.
  
The original \textsc{mercury} package 
\citep{1999MNRAS.304..793C} 
can accurately integrate planetary systems, 
including the effects of close encounters between
planets, very fast. The high 
performance is achieved because it intrinsically
conserve the angular and linear momentum of the system. 
Integrators which have this feature, are said to be
symplectic. Another property of them is that they
do not show any long-term build up of energy error, 
as conventional integrators do, although they do 
exhibit high frequency energy oscillations.
The original mercury package could not accurately 
simulate a planetary system which was inside a
binary. A method for adapting the \textsc{mercury}
package, to allow for this, was 
however presented in \citet{2002AJ....123.2884C} 
and we have here implemented those changes.

\section{The Kozai Mechanism} \label{sec:kozai}
The Kozai mechanism was first discussed in 
\cite{1962AJ.....67..591K}. He analyzed how the orbits of 
inclined asteroids in the solar system were influenced by 
Jupiter. In this case, the mass of the asteroid is much smaller 
than the mass of Jupiter. This is also true if we instead 
take the example of a Jupiter size extra-solar planet and a 
companion star and thus the same analysis holds for extra-solar 
planets within wide and inclined binaries 
\citep{1997AJ....113.1915I} (though see \citep{2002Icar..158..434C,
2003Icar..162..230C} for a 
corrected version of the Kozai equations). 
The Kozai mechanism causes the eccentricity of the planet 
to vary periodically, if the orbits of the planet and the 
companion star are sufficiently inclined. 
It has been 
used to explain the high eccentricity of the planet around 
16 Cygni B \citep{1997Natur.386..254H,1997ApJ...477L.103M} 
and also for more general studies 
of the eccentricities of planets in binary systems 
\citep{2004AIPC..713..269H,2005ApJ...627.1001T}. Its importance for more
general three-body systems has also been discussed in
\citet{2000ApJ...535..385F}. Furthermore, 
\citet{2004AJ....128.1899C} gives a detailed
discussion and analysis of the Kozai mechanism and 
use it to explain the orbits of irregular satellites in the 
solar system.

Returning to the application of the Kozai mechanism 
in this letter, one observes that it is only 
significant if the initial 
inclination, $i_0$, between the orbits of the planet 
and the companion star is greater than 39.23\degree 
($\sin^{-1}{(\sqrt{2/5})}$), which hereafter will be
referred to as the critical angle, $i_{\rm crit}$. 
If this is the case, the inclination of the planet will
oscillate between $i_{\rm crit}$ and $i_0$ and one also
observes that $\sqrt{1-e^{2}} \cos{(i)}$ is constant
\citep{1962AJ.....67..591K}. It can thus be shown,
that the maximum eccentricity, $e_{\rm max}$, 
of the planets orbit is \citep{1997AJ....113.1915I} :

\begin{equation} \label{eqn:emax}
e_{\rm max} = \sqrt{1-5/3\cos^{2}{(i_{0})}} . 
\end{equation} where it is assumed that the the 
planets initial eccentricity, $e_0$ is close to zero.
This is a lower limit on $e_{\rm max}$, since if  $e_0$
is greater than 0, the maximum eccentricity will 
increase. Using equation \ref{eqn:emax}, one can see
that if for example $i_0 = 60$\degree, 
the eccentricity will reach a maximum value of $e_{max} = 
0.76$.

It is important to note that the effects of the Kozai 
mechanism can be washed out. The effects of
non-spherical stars \citep{1980A&A....89..100S,
1997AJ....113.1915I}, relativistic precession 
\citep{1997Natur.386..254H,2000ApJ...535..385F}
and planet-planet interactions over long timescales 
\citep{1997AJ....113.1915I} can all decrease the
eccentricity build-up. If the Kozai mechanism is to 
dominate over the planet-planet interactions, its
timescale must be sufficiently short. This depends strongly 
on the semi-major axis of the companion star and the
planet and thus,
in order to see a large eccentricity 
build-up, the semi-major axis of
the companion star must not be too large. 

\begin{figure}
\resizebox{7.5truecm}{!}{\includegraphics
{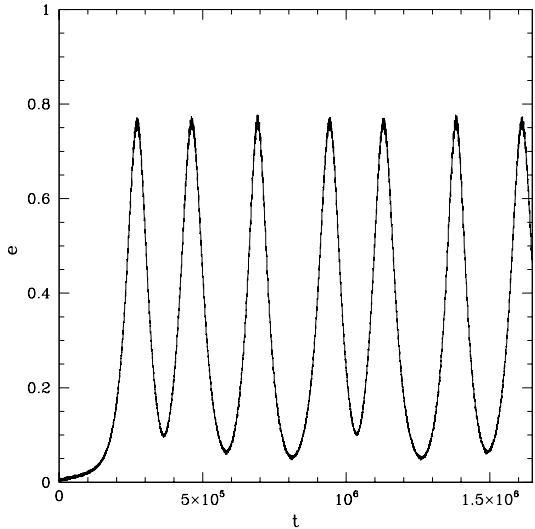}}
\caption{This plot shows the eccentricity evolution 
of a Neptune mass planet, with the same semi-major 
axis and eccentricity as Neptune ($a=30.08$ AU and
$e=0.0088$), orbiting a one solar mass star within 
a binary system, where the properties of the 
companion star are:  $M_{\rm c} = 0.5\msun$, 
$a_{\rm c} = 300$ AU, $e_{\rm c} = 0.3$ and $i_{\rm c} 
= 60$\degree.
On the horizontal axis we plot the time in years 
and on the vertical axis the eccentricity of the 
planet. The triple system is formed at $t=0$ and 
the duration of the run spans 10 000 orbital
periods of Neptune. The oscillations observed 
are in good agreement with what we expect from
the Kozai mechanism.}
\label{fig:kozai}	
\end{figure}

To illustrate the Kozai mechanism, we ran simulations 
with the \textsc{mercury} code with a 
Neptune mass planet on Neptune's orbit ($e_{\rm p} = 0.0088$ 
and $a_{\rm p} =30.07$ AU), orbiting a one solar mass star, 
with a companion star with semi-major axis $a_{c} = 300$ AU, 
eccentricity $e_{c} = 0.3$, inclination $i_0 = 60$\degree and 
mass $M_{\rm c} = 0.5 \msun$. The parameters 
of the companion star are chosen to be representative for the 
known companions to planet-hosting stars in the solar 
neighborhood \citep{2006astro.ph.10623D}.

We plot the evolution of the eccentricity of the planet in 
one of these runs in Fig. \ref{fig:kozai}. As can be seen, 
its eccentricity $e_p$ oscillates, as predicted by the Kozai 
mechanism, with maximum amplitude $e_{max} = 0.77$. We can 
compare this to the theoretical value, given by equation 
\ref{eqn:emax}, which gives that $e_{max}$ should be equal to 
0.76. Furthermore, one can see that the initial build-up of 
eccentricity is slow and then accelerates. It can be 
shown \citep{1997AJ....113.1915I} that the time to 
reach the first eccentricity maximum scales as
$T \propto a^{-3/2}_{\rm p}$, where $a_{\rm p}$ 
is the semi-major axis of 
the planet, under the assumption that the binary 
parameters and the planets eccentricity are kept constant.
We can now compare the Kozai 
time-scales of for example Uranus and Neptune and find that 
it is about twice as long for Uranus as it is for Neptune.
Thus, in a 
planetary system consisting of two or more planets, the 
eccentricity of the outer planet will reach its maximum 
when the eccentricity of the inner planets is still 
almost at its initial value. This is because the eccentricity 
stays at its original value for quite a long time until 
the growth starts, but once it does, it is very fast, 
as can be seen in Fig. \ref{fig:kozai}.

Thus, if an extra-solar planetary-system is within 
a highly inclined and not too wide binary, the Kozai 
mechanism can lead to a very large increase in the 
eccentricity of the outer planet. When this occurs, 
the inner planets will still be almost unaffected 
by the companion star. If the inclination is high 
enough, the eccentricity of the outer planet will 
be large enough for the orbits of the two outer 
planets to cross and hence strong planet-planet 
encounters are likely to occur. It is however very 
important to note, that when the eccentricity of the 
outer planet is at its maximum, the inclination between 
it and the companion star is at its minimum. This also 
means that the outer planet is inclined to the rest of 
the planets. This decreases the likelihood of 
planet-planet interactions. The outcome is that it 
can take several Kozai cycles before the planets 
have strong encounters (which depends on
the phasing of the planets). Once they occur, such 
strong encounters can lead to large changes in both 
the semi-major axis and the eccentricity of the 
planets, or even the ejection of 
one or two of them, as we show in the following 
section.

\section{Numerical simulations - Results} \label{sec:runs}
\begin{figure}
\resizebox{7.5truecm}{!}{\includegraphics{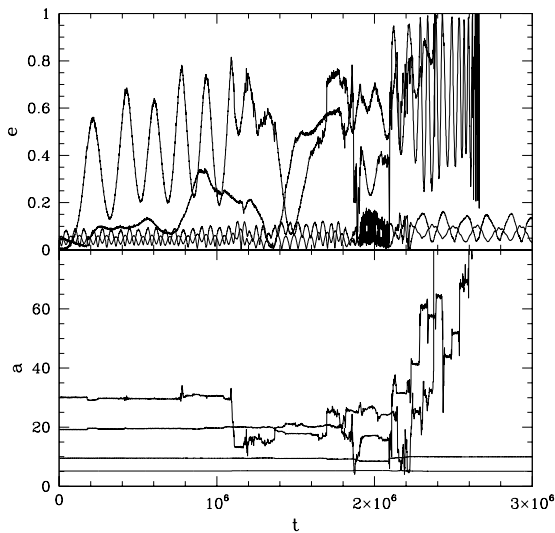}}
\caption{This plot shows the evolution of the the 
four gas giants in our solar system, placed within 
a binary system. The properties of the companion 
star are: $a_{\rm c} = 300$ AU, $e_{\rm c} = 0.3$,
$m_{\rm c} = 0.5 \msun$ and $i_{\rm c} = 60$\degree. 
The orbital elements of the companion star does 
not change during the duration of the simulation. 
On the horizontal axis we have plotted the time 
in years and on the vertical axis we have in the 
lower figure plotted the semi-major axis in AU 
and in the upper figure the eccentricity of 
the planets.}
\label{fig:FullRun}
\end{figure}

As shown in the previous section, the Kozai mechanism 
can lead to strong planet-planet interactions. We 
investigate this here numerically, by taking the 
four gas giants of our own solar system around a one 
solar mass star and evolve the system within a binary. 
In Fig. \ref{fig:FullRun} we plot the results of 
such a run, where the system has a companion star with 
$m_{\rm c} = 0.5\msun$, $a_{\rm c} = 300$ AU, $e_{\rm c} = 
0.3$ and $i_{\rm c} = 60$\degree. 

As can be seen, there is an almost immediate build-up 
in the eccentricity of the outer planet, and after only 
about 3 million years the two outermost planets have been ejected, 
due to strong planet-planet interactions. By looking at 
the history of the close encounters that the two outer 
planets in the run in Fig. \ref{fig:FullRun} had, 
it is possible to better understand what happens.

Both of them have a series of close encounters with the 
other planets and themselves, which coincide with their 
periods of high eccentricity over the first million years 
of the run. After this, there is a large change in the 
semi-major axis of the outermost planet, which leads to 
even more strong encounters between all the planets. Both 
the outer planets are then ejected shortly after each other,
at around $2.5 \times 10^6$ years, due to strong encounters 
with the innermost planet. One can from the close encounter
history clearly see that all the sudden changes in the orbital
elements of the planets seen in Fig. \ref{fig:FullRun} are solely
due to close encounters with other planets.

It could be possible that the results seen in Fig. 
\ref{fig:FullRun} are due to the initial phasing of 
the planets and that planetary systems normally are 
less sensitive. To test this, we performed 10 different 
realizations of the run in Fig. \ref{fig:FullRun}, where 
we varied the initial phasing of the planets. The results 
of these runs were similar to those of the run shown
in Fig. \ref{fig:FullRun}. In all of the 10 runs performed, 
one or two outer planets in the system were ejected 
within $6 \times 10^6$ years. It is important to note, 
that in all the 10 runs, the companion star suffers no 
close encounter with any of the planets, hence the ejection 
of planets are not due to strong interactions 
between the planets and the star.

To further test our hypothesis that the instabilities seen 
in Fig. \ref{fig:FullRun} are caused by the
Kozai mechanism, through the increased eccentricity 
of the outer planet and the following planet-planet
interactions,
we then performed another set of 10 
runs. These had the same initial conditions 
for the companion star as in Fig. \ref{fig:FullRun}, 
but with the initial inclination between the planets 
and the companion star equal to 10\degree. Hence, 
the initial inclination between the orbital plane 
of the binary and that of the planets is much less 
than the critical angle (39.23\degree), and so the
Kozai mechanism should not be active. In all the 
10 runs (which lasted for $10^8$ years), there is 
no build-up of eccentricity for the outer planet 
and thus we see no strong planet-planet encounters, 
and no ejections, as is expected from the 
Kozai theory.

If it is only planet-planet interactions that 
cause the ejection of one or more planets in 
the runs we performed, the outcome should be 
similar if we remove the companion star once 
the eccentricity of the outer planet is high. 
To test this, we removed the companion star 
at a time when the eccentricity is at its 
maximum in the 10 highly-inclined runs 
discussed above. As expected, we still see 
planet ejections within the same time-scales, 
although the strong planet-planet encounters 
that lead to the ejections are not the same, 
which is also expected.

We have also performed simulations, where we 
place the four giants in  a very wide binary 
to see whether the Kozai mechanism could be 
observed. This should not be the case, since 
the Kozai timescale becomes very long in this 
case, and its effect should be suppressed by the 
precession of the planets due to planet-planet 
interactions \citep[see also][]{1997AJ....113.1915I}. 
We again used the giant planets of the solar 
system as our test object, and this time placed 
a companion star at 1000 AU on a nearly circular 
orbit with an inclination of 60\degree. The 
simulation showed that no secular evolution 
due to the effects of the companion star occurred
during the $10^8$ years that we 
ran the simulation for.

\section{Discussion} \label{sec:discussion}
As we have seen, a planetary system can be strongly 
affected by the presence of a companion star, even if 
the semi-major axis of the companions orbit is large. 
However, this is only true if the initial inclination 
between the orbital planes of the planets and the companion 
star is larger than the critical angle, 39.23\degree. 
If we assume that all planetary systems that are in binaries 
also were formed in them, it would be unlikely to 
find systems with such a high mutual inclination. This is
because one would expect the rotational axis of the stars to 
be aligned and hence the disk out of which the planets
form would be in the same plane as the companion star.
However, if the planetary system is formed around a single
star, which is later exchanged into a binary, the orientation 
of the planets orbital planes will be random. One can show, 
that for systems who have formed this way, 77\%  will have 
an initial inclination above 39.23\degree and hence be 
in the region where the Kozai mechanism can be important.
To understand how important the Kozai mechanism is for 
extra-solar planets, it is thus necessary to know how 
often stars are exchanged into binaries 
\citep{2005ApJ...633L.141P,2006ApJ...641..504A,
2006astro.ph..6645P}. Stars are formed in groups.
Thus, early on, stellar encounters will be frequent in such
crowded environments. This means that it might be 
quite common for stars to be exchanged in and out of 
binaries. It is therefore interesting 
to study the fraction of single stars which have never
been in a binary or even suffered a close encounter with 
another star. We term such stars \textit{singletons}. 
The fraction of single stars has been calculated
\citep{Malmberg2006B} and the results show that 
a substantial fraction of the stars are not 
singletons.

To understand how important such interactions between 
stars are for planetary systems, we need a criterion
for when they can cause significant damage to them. 
In the case that the star is exchanged into a binary for 
several $10^5$ years, a companion star can cause several 
planets to be ejected, via planet-planet interactions
induced by the Kozai mechanism.  In the example of the solar 
system, Neptune's and Uranus orbits will cross if the 
inclination between the companion star and the 
plane of the planets is greater than 43\degree. This
will be the case in 73\% of the binary systems formed in 
stellar encounters. It is thus
possible that planetary systems initially resembling our
own solar system may more closely resemble the 
observed extra-solar planetary systems due to
planet-planet scattering induced by the Kozai mechanism. 
This is true also for single stars with planetary systems, 
since these may have previously been in a binary and later
exchanged out of it.

If it
is common for young planetary systems to be stripped
of one or more planets while they are in young stellar 
clusters, we should be able to observe some of these
planets. It is interesting to note that free-floating
planetary mass objects has been observed in dense stellar
clusters \citep[see for example][]{2000MNRAS.314..858L}. 
These objects might be young planets which have been ejected
from its host system, but may also have been formed in 
other ways \citep[eg.][]{2004A&A...427..299W} 

From the simulations which  we have performed, we do 
not see the formation of so-called hot Jupiters. 
These are Jupiter mass planets found at very small 
distances (a few stellar radii) from the host star, 
on almost circular orbits. However, hot Jupiters 
may be formed from planets that have strong tidal 
encounters with the host star, which however requires 
that the planets have very eccentric orbits 
\citep{2005Icar..175..248F}. If the inclination
between the planets and the companion star is near 
90\degree the Kozai mechanism will, in the absence
of relativistic effects, cause such high
eccentricities, if the binary is not too wide. 
Thus, it might be possible to 
explain hot Jupiters in this way. 
However, observations imply that at least some highly
eccentric planetary systems have avoided tidal
circulation \citep{2003ApJ...589..605W}.
We will return to this problem in a subsequent paper.

As we have seen above, no strong secular effects occur
in our runs with planetary systems within a  
co-planar binary. This is true as long as the separation 
between the companion star and the central star is large. 
However, if the separation is smaller than 100 AU, 
one or more planets might be ejected from the system 
\citep{2005ApJ...618..502M}. Here, we do not see any 
significant build-up of the planets 
eccentricities in the low inclination case. However, 
we do see a small increase in the mutual inclinations
between the planets and their respective eccentricities.

\section{Summary} \label{sec:summary}
We have simulated the evolution of 
planetary systems, resembling 
our own solar system, inside wide binaries. 
We find, that if the inclination between
the orbital plane of the companion star 
and the orbital planes of the planets is
larger than 39.23\degree, the Kozai
mechanism will lead to an increase in
the eccentricity of the outer planet, if
the binary is not to wide. The increased
eccentricity of the outer planet leads to 
strong planet-planet interactions in the system,
which can lead to the ejection of one or more
planets and also that the remaining planets
are left on more eccentric orbits than before. As
stars are formed in groups, they
will, during the first 100 million years after their
birth, be in a crowded environment. This means that they
can exchange into binaries in stellar encounters, with a 
random orientation of the orbital plane of the
planets, with respect to the orbital plane of the 
companion star. Thus, a planetary system resembling 
our own solar system formed around a single star,
is at risk of exchanging into a binary and being 
stripped of one or more planets.

\section*{Acknowledgments}
MBD is a Royal Swedish Academy Research Fellow supported by a grant
from the Knut and Alice Wallenberg Foundation. The simulations 
performed in this letter were done on computer 
hardware which was purchased with grants from the the Royal 
Physiographic Society in Lund. We thank the anonymous referee
for useful comments.

\bibliography{planetrefs}
\bibliographystyle{mn2e} 
\label{lastpage}

\end{document}